\documentclass[12pt,a4paper,dvipdfmx]{article}
\usepackage{graphicx}
\usepackage{color}
\usepackage{bm}
\usepackage{epsfig}
\usepackage{amsmath,amssymb}
\usepackage{booktabs}
\makeatletter

\begin{document}

\title{Constraints of chromoelectric dipole moments to natural SUSY type sfermion spectrum}
\author{
  \centerline{
    Nobuhiro~Maekawa$^{1,2}$\footnote{maekawa@eken.phys.nagoya-u.ac.jp}
    ,~ 
    Yu~Muramatsu$^{3}$\footnote{yumura@mail.ccnu.edu.cn}
    ~and 
    Yoshihiro~Shigekami$^1$\footnote{sigekami@eken.phys.nagoya-u.ac.jp}}
  \\[25pt]
  \centerline{
    \begin{minipage}{\linewidth}
      \begin{center}
        $^1${\it \normalsize Department of Physics, Nagoya University, Nagoya 464-8602, Japan }\\[3pt]
        $^2${\it \normalsize Kobayashi Maskawa Institute, Nagoya University, Nagoya 464-8602, Japan }\\[3pt]
$^3${\it \normalsize        Institute~of~Particle~Physics~and~Key~Laboratory~of~Quark~and~Lepton~
Physics~(MOE), Central~China~Normal~University,~Wuhan,~Hubei~430079,
~People's~Republic~of~China}
      \end{center}
  \end{minipage}}
  \\[50pt]}
\date{}
\maketitle

\begin{abstract}
  We investigate the lower bounds of sfermion masses from the constraints of chromoelectric dipole moments (CEDMs) in the natural SUSY-type sfermion mass spectrum, in which 
stop mass $m_{\tilde t}$ is much smaller than the other sfermion masses $m_0$.
The natural SUSY-type sfermion mass spectrum has been studied since 
the supersymmetric (SUSY) flavor-changing neutral currents (FCNC) are suppressed because of large
sfermion masses of the first two generations, and the weak scale is stabilized because of the light stop.
  However, this type of sfermion mass spectrum is severely constrained by CEDM,
because the light stop contributions to the up quark CEDM are not decoupled in the limit 
$m_0\rightarrow\infty$, while the down quark CEDM is decoupled in the limit.
It is important that the constraints are severe even if SUSY-breaking parameters (and Higgsino mass)
are taken to be real because complex diagonalizing matrices of Yukawa matrices, which are from complex 
Yukawa couplings, generate nonvanishing CP phases in off-diagonal elements of sfermion mass matrices.
We calculate the CEDM of up and down quarks numerically in the minimal
SUSY standard model, and give the lower bounds for stop mass and 
the other sfermion masses.
We show that the lower bound of stop mass becomes 7 TeV to satisfy the CEDM constraints
from Hg EDM.
The result is not acceptable if the weak scale stability is considered seriously. 
We show that if the up-type Yukawa couplings are taken to be real at the grand unification scale, 
the CEDM constraints are satisfied even if $m_{\tilde t}\sim 1$ TeV.

\end{abstract}

\newpage

\section{INTRODUCTION}
\label{sec:intro}

The supersymmetric (SUSY) extended standard model (SM) is one of the most promising candidates as the model beyond the SM. 
The minimal SUSY SM (MSSM) can realize the stability of the weak scale
and  provide a
 candidate of the dark matter. Moreover, it is consistent with 
the grand unified theory (GUT)\cite{GG} since the three gauge couplings meet at a
scale which is called the GUT scale. 

However, the supersymmetry must be broken because the supersymmetric 
partners of the quark
and lepton, which are the squark and slepton, have not been found yet, and the
scale is expected to be around the weak scale, not to destabilize the weak scale.
Generically, the SUSY-breaking parameters which violate the flavor and CP symmetry induce too large flavor-changing neutral current (FCNC) processes (the SUSY flavor problem) and the CP-violating processes (SUSY
CP problem). 
To avoid these problems,
we usually assume the universal sfermion masses  and/or real
SUSY-breaking parameters at a scale.
One more interesting possibility is the decoupling solution in which 
the SUSY-breaking scale is taken to be much higher than the weak scale. 
Then the SUSY contributions to the FCNC processes and CP-violating 
processes are suppressed by decoupling features.
The sfermion masses are required to be $O(1000)$ {\rm TeV} to sufficiently
suppress the contribution to $\epsilon_K$ if  off-diagonal elements of 
sfermion mass matrices are not suppressed\cite{masiero}.
This possibility  has been
examined more in detail since the observed Higgs mass 125 GeV
\cite{Higgs} 
requires a higher SUSY-breaking scale\cite{HighScale}. 
Unfortunately, such higher SUSY-breaking parameters result in the destabilization of the weak scale; i.e., strong fine-tuning is needed to obtain
the weak scale. 

One possible way to improve the fine-tuning is to make the
stop mass lower, around 1 TeV, while the other scalar fermions (sfermions) 
have much 
larger masses than 1 TeV to suppress the FCNC and CP-violating 
processes. Such sfermion mass spectrum is called effective SUSY-type
sfermion masses or natural SUSY\cite{naturalSUSY}.  Unfortunately, it has been pointed out
that the sufficiently large sfermion masses are difficult to be taken
since large sfermion masses and small stop mass at the GUT scale tend to 
result in negative stop mass square at the weak scale via two loop
renormalization group effects. Large gluino mass can improve the situation,
because it contributes to the stop mass square positively. Roughly, 
squark masses except stop mass must be smaller than 5 times gluino mass.
Therefore, if the gluino mass is around 2-3 TeV, which is target of LHC
experiment, $O(10)$ {\rm TeV} is the upper limit. Then, the off-diagonal
elements of sfermion mass matrices must be suppressed.
One way to suppress the off-diagonal elements is to require the modified 
sfermion universality in which all sfermion masses except third generation 
${\bf 10}$ dimensional fields of $SU(5)$ are universal at the GUT scale\cite{E6Fam, Muniversal} as
\begin{equation}
\tilde m_{\bf 10}^2=\left(\begin{array}{ccc}
                               m_0^2 & 0 & 0 \cr
                              0  & m_0^2 & 0 \cr
                              0  & 0 & m_3^2
                          \end{array}\right), \quad
\tilde m_{\bf \bar 5}^2=\left(\begin{array}{ccc}
                               m_0^2 & 0 & 0 \cr
                              0  & m_0^2 & 0 \cr
                              0  & 0 & m_0^2
                          \end{array}\right).
\end{equation}
Such mass spectrum can be naturally obtained in $E_6$ GUT\cite{E6, E6neutrino, E6matter} with family

symmetry\cite{E6Fam}. 
Here the universality for all $\bf\bar 5$ fields of $SU(5)$ is important to obtain sufficiently small FCNC processes even if the diagonalizing matrices
for $\bf\bar 5$ fields have large mixings. Therefore,
when the diagonalizing matrices of ${\bf 10}$ fields and 
${\bf\bar 5}$ fields of $SU(5)$ are estimated by 
Cabibbo-Kobayashi-Maskawa (CKM) matrix $V_{CKM}$ and Maki-Nakagawa-Sakata (MNS) matrix $V_{MNS}$ as
\begin{equation}
V_{\bf 10}\sim V_{CKM}\sim \left(\begin{array}{ccc}
                                          1 & \lambda & \lambda^3 \cr
                                          \lambda & 1 & \lambda^2 \cr
                                          \lambda^3 & \lambda^2 & 1
                                          \end{array}\right),\quad
V_{\bf \bar 5}\sim V_{MNS}\sim \left(\begin{array}{ccc}
                                          1 & \sqrt{\lambda} & \lambda \cr
                                          \sqrt{\lambda} & 1 & \sqrt{\lambda} \cr
                                          \lambda & \sqrt{\lambda} & 1
                                          \end{array}\right),                                          
\end{equation}   
which are expected in some GUT models\cite{E6matter, naturalGUT}, 
off-diagonal elements of sfermion mass matrices can be suppressed as
\begin{equation}
\tilde m_{\bf 10}^2=V_{\bf 10}^\dagger \left(\begin{array}{ccc}
                                                                 m_0^2 & 0 & 0 \cr
                                                                 0  & m_0^2 & 0 \cr
                                                                 0  & 0 & m_3^2
                                                          \end{array}\right)V_{\bf 10}
                        =m_0^2{\bf 1}
                        +(m_3^2-m_0^2)
\left(\begin{array}{ccc}
\lambda^6 & \lambda^5 & \lambda^3 \cr
\lambda^5 & \lambda^4 & \lambda^2 \cr
\lambda^3 & \lambda^2 & 1
  \end{array}\right)                                          
\end{equation}                                                    
and most of flavor and CP constraints can be satisfied.
Lepton flavor violation processes like $\mu\rightarrow e\gamma$ or
$\tau\rightarrow \mu \gamma$ may be seen\cite{FCNC} if 
$m_3\sim O(100)$ GeV, but unfortunately, we lost the strong reason
to take $m_3\sim O(100)$ GeV after discovery of Higgs particle 
because  lower bound of stop masses becomes 
around 1 TeV in the MSSM in order to realize the Higgs mass $\sim 125$
GeV.
Here $\lambda\sim 0.22$ is the Cabibbo mixing angle.

However, even if this modified universal sfermion mass spectrum with real
SUSY-breaking parameters are adopted, 
the EDM  constraints from 
the experimental bound\cite{EDMexp} as
\begin{eqnarray}
d_N&<&3.0\times 10^{-26}\: e\:{\rm cm} \label{NEDM}\\
d_{Hg}&<&7.4\times 10^{-30}\: e\:{\rm cm} \label{HgEDM}
\end{eqnarray}
become severe. The essential point is that the sfermion mass matrices of
$\bf 10$ fields of $SU(5)$ in super-CKM basis where quark and lepton 
mass matrices
 are diagonal have complex off-diagonal elements generically unless sfermion masses are universal because
 Yukawa couplings are complex, and therefore, diagonalizing matrices are 
 complex to obtain
 the Kobayashi-Maskawa (KM) phase. Therefore,  the diagram with the complex
 off-diagonal elements can contribute to the (chromo) EDM, and give
 severe constraints to the off-diagonal elements\cite{HisanoShimizu}.
 Most of contributions to (chromo) EDM are decoupled in the limit
 $m_0\rightarrow \infty$ with finite $m_3$, but some contributions
 to up quark (chromo) EDM are not decoupled in the limit\cite{Ishiduki}.
The constraint becomes
 \begin{equation}
 {\rm Im}\left[\frac{({\tilde m}_{\bf 10}^2)_{31}}{m_0^2}\frac{({\tilde m}_{\bf 10}^2)_{13}}{m_0^2}\right]<\left\{\begin{array}{l}
 5.3\times 10^{-6 } ({\rm Hg}) \\
 1.6\times 10^{-4} ({\rm neutron}) 
 \end{array}\right\} 
 \left(\frac{m_3}{2{\rm TeV}}\right)^2, 
 \end{equation}
 which are obtained by the mass insertion approximation (MIA)\cite{MIA}
 with certain mass spectrum of SUSY particles.
Here we have used the relations between neutron (Hg) EDMs and CEDM of quarks in Ref. \cite{Hisano}(\cite{Pich}) for neutron (Hg) EDM.  Although the ambiguity in theoretical calculation of EDMs  is large especially for Hg\cite{Pich, Engel}, 
we give the constraints to the model by neglecting the uncertainty in this paper.
Since $({\tilde m}_{\bf 10}^2)_{31}/{m_0^2}\sim({\tilde m}_{\bf 10}^2)_{13}/m_0^2\sim\lambda^3$, 
the predicted EDM of Hg becomes about 20 times larger than the experimental bound even if we take
 stop mass $m_{\tilde t}\sim 2$ TeV and the diagonalizing matrices of 
 $V_{\bf 10}$ have small mixings like CKM matrix.
This severe constraint looks to be general for almost all models with natural
SUSY spectrum, and it is important to study the solution to this problem
if natural SUSY spectrum is studied. Note that this problem cannot be
solved by real SUSY-breaking parameters, because the CP phases of
off-diagonal elements of sfermion mass matrices come from the complex
Yukawa couplings which are usually assumed to obtain nonvanishing KM
phase. We call this (chromo) EDM problem new SUSY CP problem in this 
paper.

In this paper, we focus on the new SUSY CP problem in the models with 
natural SUSY spectrum. One easy solution is to take large stop mass
although large stop mass destabilizes the weak scale. 
Another one is to take diagonal up-type Yukawa matrix and therefore,
diagonalizing matrix of up-type quark becomes unit matrix, although
it is not so easy to obtain it in natural way.
One more interesting solution to this new SUSY CP problem is to take
real up-type Yukawa couplings to suppress the (chromo) EDM and complex
down-type Yukawa couplings to obtain the KM phase. 
That possibility has been 
proposed and discussed in the $E_6$ GUT with family and CP symmetry\cite{Ishiduki, E6CP},
which is spontaneously
broken by a CP phase in Higgs vacuum expectation value (VEV) which
breaks the family symmetry also. 

In this paper, we calculate the chromo EDMs (CEDMs) of up and down 
quarks in the MSSM with
different boundary conditions because the CEDMs give more severe constraints than the usual EDMs for natural SUSY-type models with real SUSY-breaking parameters.
(In the recent paper\cite{Nakai}, the EDM constraints in natural SUSY models with complex SUSY-breaking
parameters has been discussed. However, the contributions discussed in this paper often give stronger 
constraints to the natural SUSY-type models even if SUSY-breaking parameters are real.)
We also discuss the decoupling features of those constraints.
First, we show that the nondecoupling feature of the up-quark CEDM
contribution by the MIA and that the stop mass must be larger than
10 TeV to satisfy the CEDM constraints. 
Second, we calculate the up- and down-quark CEDMs numerically in the MSSM
 and
show that the stop mass and the other squark masses  must be larger than 7 TeV to satisfy the
up- and down-quark CEDMs. If the real up-type Yukawa couplings are adopted, 
CEDM constraints can be satisfied even if the stop mass is $O(1)$ TeV 
although the other squark masses must be larger than 7 TeV.
Finally, we discuss the predictions. 

\section{ROUGH ESTIMATE OF CEDM}
\label{sec:CEDM}
In this section, we calculate CEDM of up quark in the modified universal 
sfermion mass spectrum by MIA, and see 
the nondecoupling feature in the limit $m_0\rightarrow \infty$ when 
$m_3$ and the other SUSY-breaking parameters are fixed. 

\begin{figure}[h]
  \begin{center}
    {\epsfig{figure=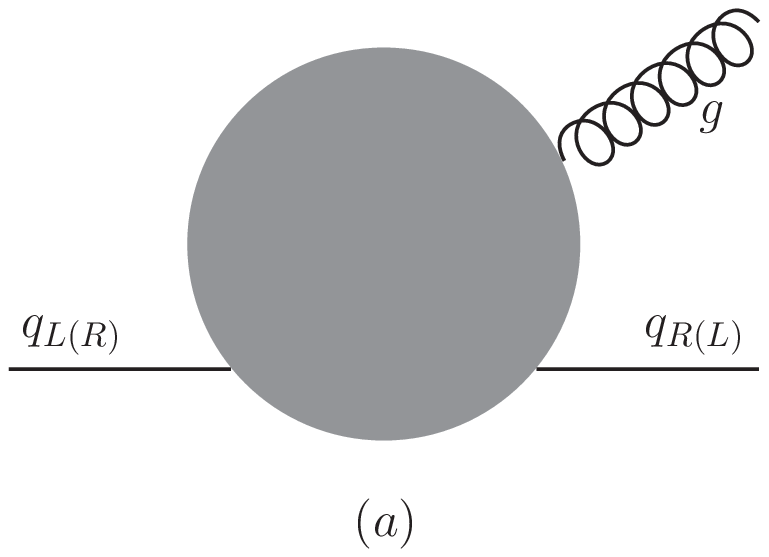,width=0.4\textwidth}}\hspace{0.3em}{\epsfig{figure=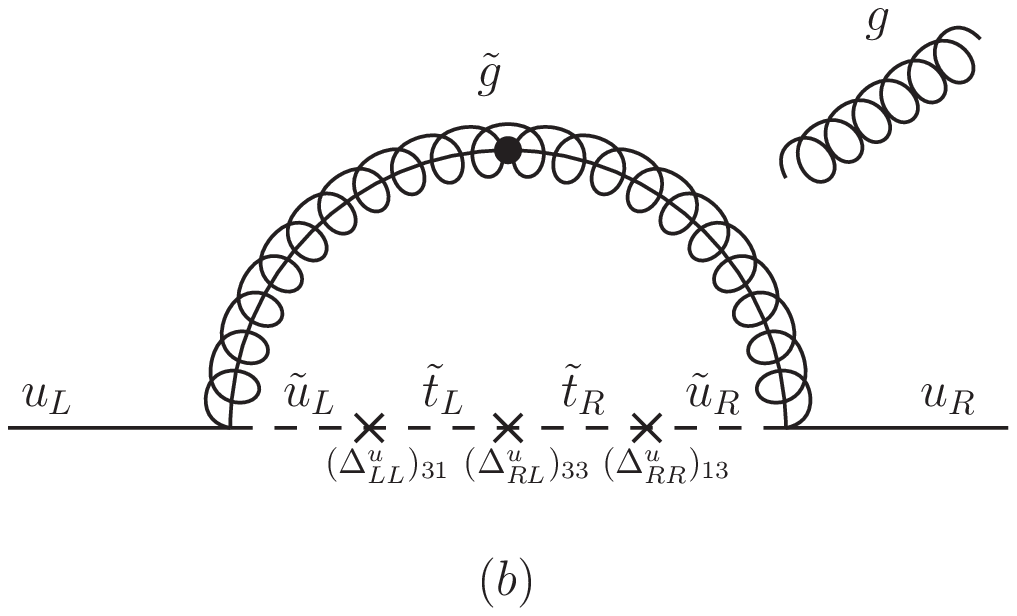,width=0.5\textwidth}}
    \caption{ Diagrams contribute to $d_q^C$ (a) and $d_u^C$ (b). $(\Delta_{AB}^u)_{ij} \, (A, B=L\, {\rm or}\, R,\, i,j=1,2,3)$ is the element of $6\times 6$ sfermion mass matrix [see Eq. (\ref{eq:66sfmass})].}
    \label{fig:CEDMdiagram}
  \end{center}
\end{figure}

The effective lagrangian for CEDM is described as
\begin{eqnarray}
  \mathcal{L}_{\rm CEDM}=-\frac{ig_s}{2} d_q^C \overline{q}  (G \cdot \sigma) \gamma_5 q,
  \label{eq:LCEDM}
\end{eqnarray}
where $g_s$ is the QCD coupling, $G \cdot \sigma = G^A_{\mu\nu} T^A \sigma^{\mu\nu}$, $G^A_{\mu\nu}$ is field strength of gluon, $T^A$ $(A=1,2,\cdots, 8)$ are $SU(3)$ generators and $\sigma^{\mu\nu}=\frac{i}{2}[\gamma^{\mu},\gamma^{\nu}]$. $d_q^C$ shows quark CEDM and one can calculate this by computing diagrams shown in Fig. \ref{fig:CEDMdiagram}(a). In particular, $d_q^C$ is dominated by gluino contributions in the SUSY model. In the natural SUSY-type models, the diagram in Fig. \ref{fig:CEDMdiagram}(b) dominantly contributes to 
$d_u^C$\cite{HisanoShimizu, Ishiduki}.
We estimate the magnitude of $d_u^C$ by using the diagram in Fig. \ref{fig:CEDMdiagram}(b) by MIA
as
\begin{eqnarray}
  d_u^C &\simeq& \frac{\alpha_s}{4\pi}F\left(\frac{M^2_{\tilde g}}{m^2_{\tilde u}},\frac{m^2_{\tilde t}}{m^2_{\tilde u}}\right) {\rm Im}\left[ \frac{M_{\tilde{g}}}{m_{\tilde{t}}^2} \frac{(\Delta_{RL}^u)_{33}}{m_{\tilde{t}}^2} \frac{(\Delta_{LL}^u)_{31}}{m_{\tilde{u}}^2} \frac{(\Delta_{RR}^u)_{13}}{m_{\tilde{u}}^2}\right]\nonumber \\
  &\sim& \frac{\alpha_s}{4\pi} F\left(\frac{M^2_{\tilde g}}{m^2_{\tilde u}},\frac{m^2_{\tilde t}}{m^2_{\tilde u}}\right)\frac{M_{\tilde g}A_{u33}v_u}{m^4_{\tilde{t}}}  {\rm Im}\left[ (\delta_{LL}^u)_{31} (\delta_{RR}^u)_{13}\right],
  \label{eq:duCapp}
\end{eqnarray}
where $(\delta_{AB}^u)_{ij}$ are the mass insertion parameters, defined as
\begin{eqnarray}
  (\delta_{AB}^u)_{ij} \equiv \frac{(\Delta_{AB}^u)_{ij}}{m_{\tilde{u}}^2},\,\,(A, B=L\, {\rm or}\, R,\, i,j=1,2,3),
\end{eqnarray}
and $F(x,y)$ is a loop function.
In this definition, $(\Delta_{AB}^u)_{ij}$ is an element of the $6\times 6$ sfermion mass matrix in the super-CKM basis,
\begin{eqnarray}
  M_{\tilde{u}}^2=\left(
  \begin{array}{cc}
    L_u^{\dagger} (m_Q^2 + v_u^2 Y_u^{\ast} Y_u^T) L_u & L_u^{\dagger} (v_u A_u^{\ast} - \mu v_d Y_u^{\ast}) R_u^{\ast} \\
    R_u^T (v_u A_u^T - \mu v_d Y_u^T) L_u & R_u^T (m_u^2 + v_u^2 Y_u^T Y_u^{\ast}) R_u^{\ast} \\
  \end{array}
  \right) \equiv \left(
  \begin{array}{cc}
    (\Delta_{LL}^u) & (\Delta_{LR}^u) \\
    (\Delta_{RL}^u) & (\Delta_{RR}^u) \\
  \end{array}
  \right),
  \label{eq:66sfmass}
\end{eqnarray}
where $A_u$ is a $3 \times 3$ matrix for scalar three point vertex, 
$m_Q^2$ and $m_u^2$ are $3 \times 3$ soft SUSY-breaking mass matrices, and $v_u$ and $v_d$ are MSSM Higgs VEVs.
Here, $L_u, R_u$ are diagonalizing matrices for Yukawa coupling $Y_u$ as
$L_u^TY_u R_u=\hat Y_u$ which is a diagonal
matrix of $Y_u$.  In the last similarity in Eq. (\ref{eq:duCapp}),
 we have assumed that the gluino mass $M_{\tilde{g}}$ and $(\Delta_{RL}^u)_{33}\sim A_{u33}v_u$ are real. 
Even if all SUSY-breaking parameters and Higgsino mass $\mu$ are set to be real,
$M_{\tilde u}^2$ becomes complex generically because Yukawa couplings are
taken to be complex to obtain the  KM phase and therefore diagonalizing 
matrices $L_u$ and $R_u$ are complex.
Quantitative constraint for mass insertion parameter in Eq. (\ref{eq:duCapp}) comes from the current CEDM bound, 
$|d_u^C| < 3.4\times 10^{-27} (1.0\times 10^{-25})\: {\rm cm}$, which are obtained from the present upper limit of $d_{\rm Hg} (d_{\rm N})$ in Eq. (\ref{HgEDM}) [in Eq. (\ref{NEDM})], and
the relation $d_{\rm Hg}\sim 2.2\times 10^{-3}e(d_u^C-d_d^C)$\cite{Pich}\footnote{
Here, we use the bound for $|d_u^C-d_d^C|$ as the bound for $d_u^C$. This is justified in the limit $m_0\rightarrow \infty$ because
$d_d^C$ is vanishing in the limit. In this paper, we just use the bound for $d_u^C-d_d^C$ as the bound for $d_u^C$ and $d_d^C$ even
with finite $m_0$.
}
($d_{\rm N}\sim -0.3 e(d_u^C-d_d^C)$\cite{Hisano})
, as 
\begin{eqnarray}
  {\rm Im}\left[ (\delta_{LL}^u)_{31} (\delta_{RR}^u)_{13} \right] < 
  \left\{\begin{array}{l}5.3\times 10^{-6} ({\rm Hg}) \\
                            1.6\times 10^{-4} ({\rm neutron})
                            \end{array}\right\}\left(\frac{m_{\tilde{t}}}{2 \, {\rm TeV}}\right)^2,
  \label{eq:MIconst}
\end{eqnarray}
where we have used $g_s \simeq 1$, $m_{\tilde{t}} \sim A_{u33}\sim 2$ TeV, 
$M_{\tilde{g}}\sim 1.5m_{\tilde t}\sim 3$ TeV,\footnote{
If $A$-term is smaller, the constraints become weaker, although 
$A\ll M_{\tilde g}$ usually requires a tuning in the MSSM with SUSY-breaking parameters given at the GUT scale 
because of the renormalization group effects.}
 $m_{\tilde u}\sim 10$ TeV, and the loop integral function 
$F(0.09,0.04)$ is 
$0.079$ which is obtained from 
the Appendix. When we assume that
\begin{eqnarray}
  L_u, R_u \sim V_{\rm CKM} \sim \left(
  \begin{array}{ccc}
    1 & \lambda & \lambda^3 \\
    \lambda & 1 & \lambda^2 \\
    \lambda^3 & \lambda^2 & 1 \\
  \end{array}\right)
\end{eqnarray}
and $m_{\tilde{u}} = m_{\tilde{c}} \gg m_{\tilde{t}}$, the left-hand side of Eq. (\ref{eq:MIconst}) becomes
\begin{eqnarray}
  {\rm Im}\left[ (\delta_{LL}^u)_{31} (\delta_{RR}^u)_{13}\right] \simeq \left(\frac{m_{\tilde{t}}^2 - m_{\tilde{u}}^2}{m_{\tilde{u}}^2}\right)^2\times \lambda^6 \,\,\, (\lambda=0.22).
\end{eqnarray}
So this contribution does not decouple in the limit of $m_{\tilde{u}} \to \infty$ and the size is about $10^{-4}$ that is about 20
times larger than the constraint from Hg EDM. Therefore, $m_{\tilde t} >10$ TeV is required
to satisfy the CEDM constraint in this approximation. 



On the other hand, for the CEDM of down quark, $d_d^C$, such 
contributions from flavor-violating mass insertion is decoupled when
sdown mass $m_{\tilde d}\rightarrow \infty$ because 
the right-handed sbottom mass 
$m_{\tilde b}\sim m_{\tilde d}$. Therefore, $m_{\tilde d} > 10$ TeV is expected to be required to satisfy the experimental bound of CEDM if the decoupling feature is similar to that of $d_u^C$ for
$m_{\tilde t}$.

Therefore, the CEDM of up quark is more serious  in natural SUSY
scenario.
One simple solution is that real $Y_u$ and $A_u$ are taken while $Y_d$ is 
complex that induces the KM phase. In this case, diagonalizing matrices of 
up-type quark mass matrix are also real and then $d_u^C$ is strongly suppressed. Note that $Y_u$ becomes complex through the renormalization group equation (RGE), even if $Y_u$ is real at the  GUT scale. In such a case, however, $d_u^C$ is small enough to satisfy the current experimental bound as we will show in Sec. \ref{sec:result}.

\section{EVALUATIONS AND RESULTS}
\label{sec:result}

In this section, we numerically calculate the CEDM in the MSSM 
with the modified universal sfermion mass spectrum. 

Now, we explain the procedure for the calculation of CEDMs. First of all, input parameters, which are gauge couplings $g_i$, gaugino masses $M_i$, Yukawa couplings, $A$ parameters which are couplings of three scalar's
vertex, sfermion mass matrices, and doublet Higgs masses are given at the 
GUT scale, $\Lambda_{\rm G}=2\times 10^{16}$ GeV, as
\begin{eqnarray}
&&g_1(\Lambda_{\rm G})=g_2(\Lambda_{\rm G})=g_3(\Lambda_{\rm G})=g_{\mathrm{GUT}}=0.7, \\[0.5ex]
&&M_1(\Lambda_{\rm G})=M_2(\Lambda_{\rm G})=M_3(\Lambda_{\rm G})=M_{1/2}, \label{eq:GUTboundary} \\[0.5ex]
&&Y_u\sim\left(
\begin{array}{ccc}
\lambda^6 & \lambda^5 & \lambda^3 \\
\lambda^5 & \lambda^4 & \lambda^2 \\
\lambda^3 & \lambda^2 & 1 \\
\end{array}\right),\quad
A_u\sim A_0\left(
\begin{array}{ccc}
\lambda^6 & \lambda^5 & \lambda^3 \\
\lambda^5 & \lambda^4 & \lambda^2 \\
\lambda^3 & \lambda^2 & 1 \\
\end{array}\right), \label{eq:uphierarchy} \\
&&Y_d\sim Y_e^T\sim \left(
\begin{array}{ccc}
\lambda^6 & \lambda^{5.5} & \lambda^5 \\
\lambda^5 & \lambda^{4.5} & \lambda^4 \\
\lambda^3 & \lambda^{2.5} & \lambda^2 \\
\end{array}\right),\quad
A_d\sim A_e^T\sim A_0\left(
\begin{array}{ccc}
\lambda^6 & \lambda^{5.5} & \lambda^5 \\
\lambda^5 & \lambda^{4.5} & \lambda^4 \\
\lambda^3 & \lambda^{2.5} & \lambda^2 \\
\end{array}\right), \label{eq:downhierarchy} \\
&&\tilde{m}_{\bf 10}^2=\left(
\begin{array}{ccc}
m_0^2 & 0 & 0 \\
0 & m_0^2 & 0 \\
0 & 0 & m_3^2 \\
\end{array}\right),\quad
\tilde{m}_{\overline{\bf 5}}^2=\left(
\begin{array}{ccc}
m_0^2 & 0 & 0 \\
0 & m_0^2 & 0 \\
0 & 0 & m_0^2 \\
\end{array}\right), \label{eq:MUSM} \\[0.5ex]
&&m_{H_u}^2(\Lambda_{\rm G})=m_{H_d}^2(\Lambda_{\rm G})=(500\: \mathrm{GeV})^2,
\end{eqnarray}
where $A_0$ is the typical scale of $A$ parameters, and sfermion mass matrices $\tilde{m}_{\bf 10}^2$ and $\tilde{m}_{\bar{\bf 5}}^2$ are for ${\bf 10}$ and $\bar{\bf 5}$ fields, respectively. The Higgsino mass $\mu$ is fixed by the 
value of the Z boson mass $m_Z$. ($\mu$ becomes comparatively large ($O(1)$
TeV), which may destabilize the weak scale. But we do not mind it because
large $\mu$ does not contribute much to $d_u^C$.)
Next, in order to obtain low-energy parameters from these inputs, we use two-loop RGEs based on Ref. \cite{2loopRGE}. Note that, for simplicity, we consider MSSM from GUT scale to the SUSY-breaking scale. Finally, we compute up, down and strange quark CEDMs. These CEDMs denoted as $d_q^C\: (q=u,d,s)$ are evaluated by the following one-loop formulas,
\begin{eqnarray}
&&d_u^C=c\frac{\alpha_s}{4\pi}\sum_{j=1}^6\frac{M_{\tilde{g}}}{(\hat{M}_{\tilde{u}}^2)_{jj}}\left\{\left(-\frac{1}{3}F_1(x_j^u)-3F_2(x_j^u)\right){\rm Im}[(U_{\tilde{u}}^{\dagger})_{1j}(U_{\tilde{u}})_{j4}]\right\}, \\
&&d_d^C=c\frac{\alpha_s}{4\pi}\sum_{j=1}^6\frac{M_{\tilde{g}}}{(\hat{M}_{\tilde{d}}^2)_{jj}}\left\{\left(-\frac{1}{3}F_1(x_j^d)-3F_2(x_j^d)\right){\rm Im}[(U_{\tilde{d}}^{\dagger})_{1j}(U_{\tilde{d}})_{j4}]\right\}, \\
&&d_s^C=c\frac{\alpha_s}{4\pi}\sum_{j=1}^6\frac{M_{\tilde{g}}}{(\hat{M}_{\tilde{d}}^2)_{jj}}\left\{\left(-\frac{1}{3}F_1(x_j^d)-3F_2(x_j^d)\right){\rm Im}[(U_{\tilde{d}}^{\dagger})_{2j}(U_{\tilde{d}})_{j5}]\right\},
\end{eqnarray}
where $c\sim 0.9$ is QCD correction. $\hat{M}_{\tilde{q}}^2\: (q=u,d)$ are diagonalized squark mass matrices which
are obtained as $\hat{M}_{\tilde{q}}^2=U_{\tilde{q}}M_{\tilde{q}}^2U_{\tilde{q}}^{\dagger}$, where
$M_{\tilde{q}}^2$ and $U_{\tilde{q}}$ are $6\times 6$ squark mass matrices  
and the diagonalizing matrices, respectively. 
$F_1(x)=(x^2-4x+3+2{\rm ln}x)/2(1-x)^3$ and $F_2(x)=(x^2-1-2x{\rm ln}x)/2(1-x)^3$ are coming from loop integrals and $x_j^q=\frac{M_{\tilde{g}}^2}{(\hat{M}_{\tilde{q}}^2)_{jj}}$. The current bounds \cite{EDMexp, HisanoShimizu} are
\begin{eqnarray}
&&|d_{q}^C| < 3.4\times 10^{-27}\: {\rm cm}\quad (q=u,d), ({\rm Hg}) \label{eq:boundud} \\
&&|d_{q}^C| < 1.0\times 10^{-25}\: {\rm cm}\quad (q=u,d), ({\rm neutron}) \\
&&|d_{s}^C| < 1.1\times 10^{-25}\: {\rm cm}. \label{eq:bounds}
\end{eqnarray}

We consider three types of inputs of Yukawa couplings and $A$ parameters. All of three types have the hierarchy explained above, but have different type of $\mathcal{O}(1)$ coefficients. To explain this, we show the explicit forms of these matrices:
\begin{eqnarray}
&&Y_u=\left(
\begin{array}{ccc}
y_{u11}\lambda^6 & y_{u12}\lambda^5 & y_{u13}\lambda^3 \\
y_{u21}\lambda^5 & y_{u22}\lambda^4 & y_{u23}\lambda^2 \\
y_{u31}\lambda^3 & y_{u32}\lambda^2 & y_{u33} \\
\end{array}\right),\quad
A_u=A_0\left(
\begin{array}{ccc}
a_{u11}\lambda^6 & a_{u12}\lambda^5 & a_{u13}\lambda^3 \\
a_{u21}\lambda^5 & a_{u22}\lambda^4 & a_{u23}\lambda^2 \\
a_{u31}\lambda^3 & a_{u32}\lambda^2 & 1 \\
\end{array}\right), \label{eq:YuAu} \\
&&Y_d=\left(
\begin{array}{ccc}
y_{d11}\lambda^6 & y_{d12}\lambda^{5.5} & y_{d13}\lambda^5 \\
y_{d21}\lambda^5 & y_{d22}\lambda^{4.5} & y_{d23}\lambda^4 \\
y_{d31}\lambda^3 & y_{d32}\lambda^{2.5} & y_{d33}\lambda^2 \\
\end{array}\right),\quad
A_d=A_0\left(
\begin{array}{ccc}
a_{d11}\lambda^6 & a_{d12}\lambda^{5.5} & a_{d13}\lambda^5 \\
a_{d21}\lambda^5 & a_{d22}\lambda^{4.5} & a_{d23}\lambda^4 \\
a_{d31}\lambda^3 & a_{d32}\lambda^{2.5} & a_{d33}\lambda^2 \\
\end{array}\right), \label{eq:YdAd} \\
&&Y_e=\left(
\begin{array}{ccc}
y_{e11}\lambda^6 & y_{e12}\lambda^5 & y_{e13}\lambda^3 \\
y_{e21}\lambda^{5.5} & y_{e22}\lambda^{4.5} & y_{e23}\lambda^{2.5} \\
y_{e31}\lambda^5 & y_{e32}\lambda^4 & y_{e33}\lambda^2 \\
\end{array}\right),\quad
A_e=A_0\left(
\begin{array}{ccc}
a_{e11}\lambda^6 & a_{e12}\lambda^5 & a_{e13}\lambda^3 \\
a_{e21}\lambda^{5.5} & a_{e22}\lambda^{4.5} & a_{e23}\lambda^{2.5} \\
a_{e31}\lambda^5 & a_{e32}\lambda^4 & a_{e33}\lambda^2 \\
\end{array}\right). \nonumber \\
&& \label{eq:YeAe}
\end{eqnarray}
\begin{itemize}
\item[$\bullet$] \textit{real $Y_u$ type} \\
At the GUT scale,  $y_{dij}$, $a_{dij}$, $y_{eij}$ and $a_{eij}$ are complex $\mathcal{O}(1)$ coefficients, while $y_{uij}$ and $a_{uij}$ are real $\mathcal{O}(1)$ coefficients ($i,j=1,2,3$).
\item[$\bullet$] \textit{complex $Y_u$ type} \\
  All $y_{uij}$, $a_{uij}$, $y_{dij}$, $a_{dij}$, $y_{eij}$ and $a_{eij}$ are complex $\mathcal{O}(1)$ coefficients ($i,j=1,2,3$). 
\item[$\bullet$] \textit{$E_6$ model} (with family symmetry and spontaneous CP violation) \\
  There are special relations obtained in the model in Ref.\cite{E6CP}: $y_{u11}=y_{u13}=y_{u31}=y_{e13}=y_{e21}=0$, $y_{u12}=-y_{u21}=y_{d13}=\frac{1}{3}d_q$, $y_{u23}=y_{u32}$, $y_{d23}=y_{e32}$, $y_{d33}=y_{e33}$ and $y_{e12}=-y_{e31}$. $y_{d11}$, $y_{d12}$, $y_{d22}$, $y_{d32}$, $y_{e11}$, $y_{e22}$ and $y_{e23}$ are complex $\mathcal{O}(1)$ coefficients, and $d_q$, $y_{u22}$, $y_{u23}$, $y_{u33}$, $y_{d21}$, $y_{d23}$, $y_{d31}$, $y_{d33}$ and $y_{e12}$ are real $\mathcal{O}(1)$ coefficients (there are same structures in $A$ parameters). 
\end{itemize}
For all types, we take real parameters for $M_{1/2}$, $\mu$, $A_{u33}=A_0$, and $y_{u33}=0.8$ at the GUT scale and we set $\tan\beta=7$.
In these parameters, most of the usual contributions to EDMs are strongly suppressed especially when $m_0\rightarrow\infty$.
We take $A$ parameters which have similar hierarchies to the corresponding Yukawa couplings. This situation can be realized in models
in which hierarchies of Yukawa couplings are explained by the Froggatt-Nielsen mechanism\cite{FN}. In such situation, we think it reasonable that $O(1)$ coefficients of $A$ parameters are complex number when the $O(1)$ coefficients of corresponding Yukawa couplings 
are complex  as in Ref.\cite{E6Fam, Ishiduki, E6CP}. We checked that the numerical results do not
change at all if all $O(1)$ coefficients of $A$ parameters are taken to be real for all types above.

We need to make a few comments on $\mathcal{O}(1)$ coefficient. The complex $\mathcal{O}(1)$ coefficient $C$ means that $C=|C| \exp (i\theta^{(C)})$ as $|C|$ is real $\mathcal{O}(1)$ coefficient and $\theta^{(C)}$ is random number in the ranges $0 \leq \theta^{(C)} \leq 2\pi$.  In addition, real $\mathcal{O}(1)$ coefficient means random number within the interval 0.5 to 1.5 with $+$ or $-$ signs\footnote{
We don't contain one-loop threshold corrections in the calculation because $\mathcal{O}(1)$ coefficients produce a much greater difference in the results of CEDM than the threshold correction, although we should take into account the one-loop threshold corrections when we consider two-loop RGEs.
}. 

\begin{figure}[h]
  \begin{center}
    \epsfig{figure=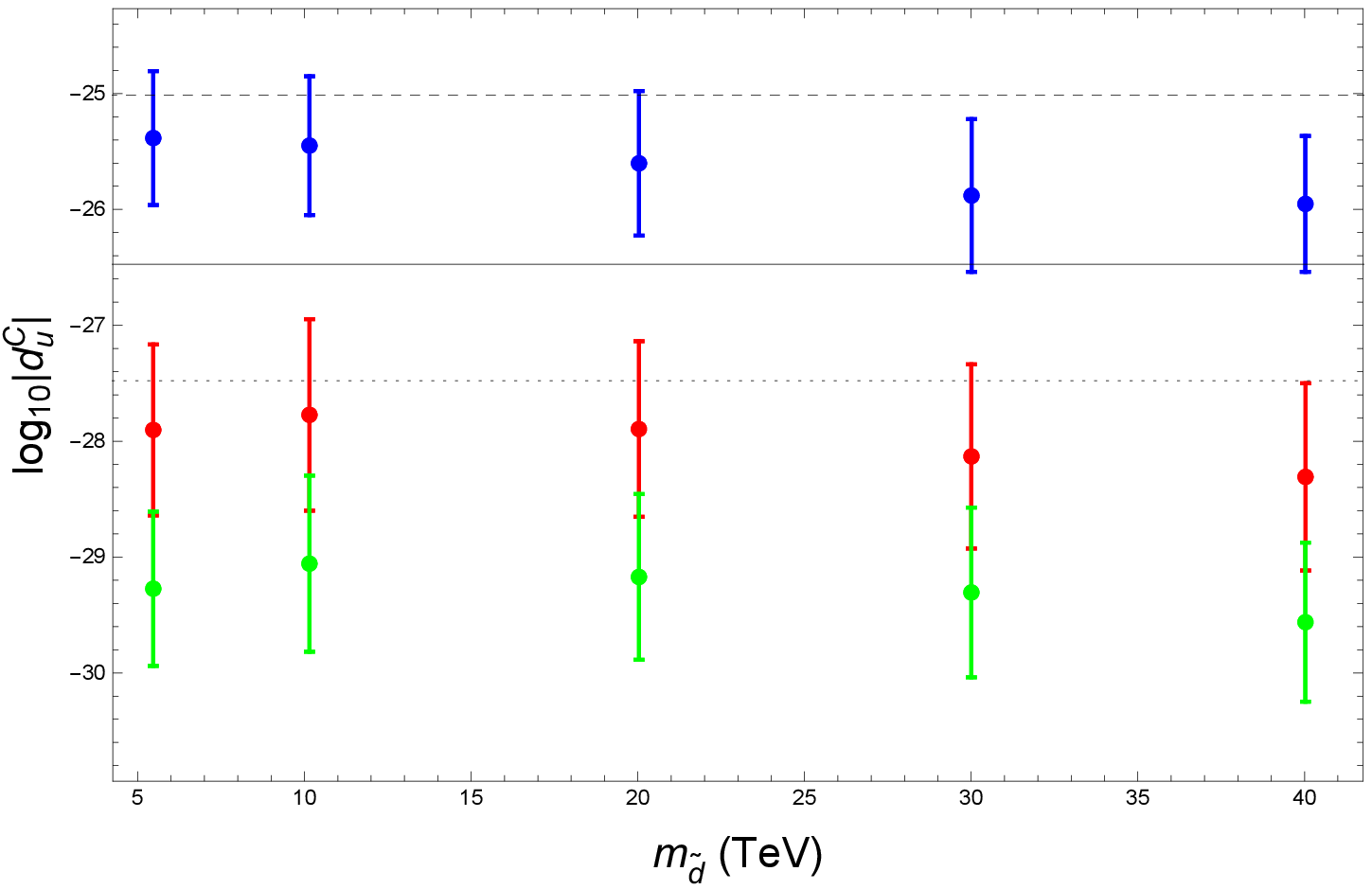,width=0.87\textwidth}\\
    \epsfig{figure=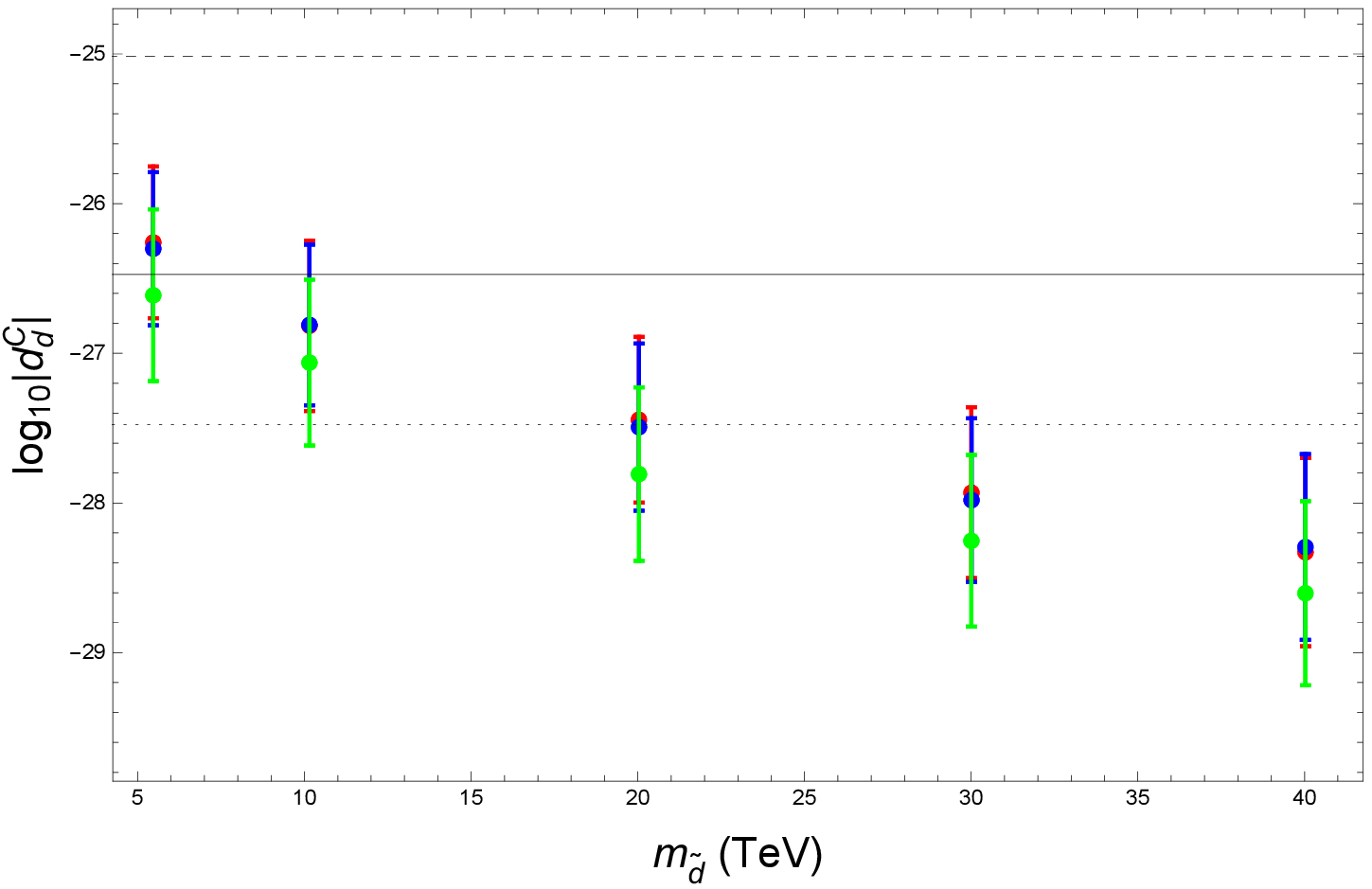,width=0.87\textwidth}
    \caption{$m_{\tilde{d}}$ dependence of $d_u^C$ (upper panel) and $d_d^C$ (lower panel). Red, Blue and Green plots are \textit{real $Y_u$ type}, \textit{complex $Y_u$ type} and \textit{$E_6$ model}, respectively. Each error bar shows the standard deviation for the predicted values of 
$\log {}_{10}|d_q^C|\: (q=u,d)$ which are obtained in various model points
with different $O(1)$ coefficients. The black solid line is the current bound from Hg EDM and the allowed region is the lower area. The dashed line shows the current bound from the neutron EDM, and the dotted line is the bound expected in future experiments of
neutron EDM. We choose $m_3$ and $M_{1/2}$ to become light stop at the SUSY scale (1 TeV), and in these figures, we set $m_{\tilde{t}}=(2000\pm 250)$ GeV. We also set $A_0=-1$ TeV.}
    \label{fig:m0depduCddC}
  \end{center}
\end{figure}

\begin{figure}[t]
  \begin{center}
    \epsfig{figure=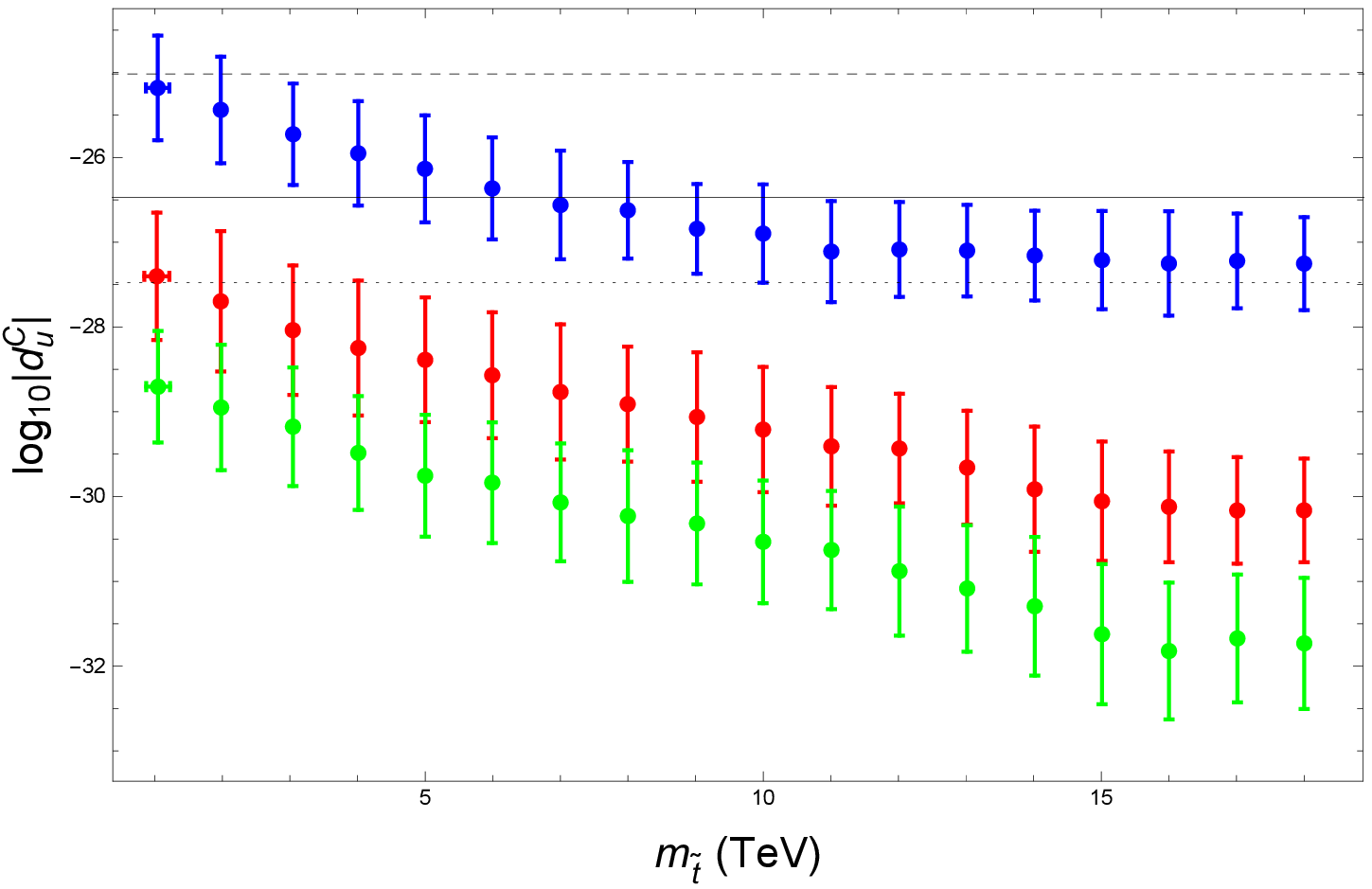,width=0.87\textwidth}\\
    \epsfig{figure=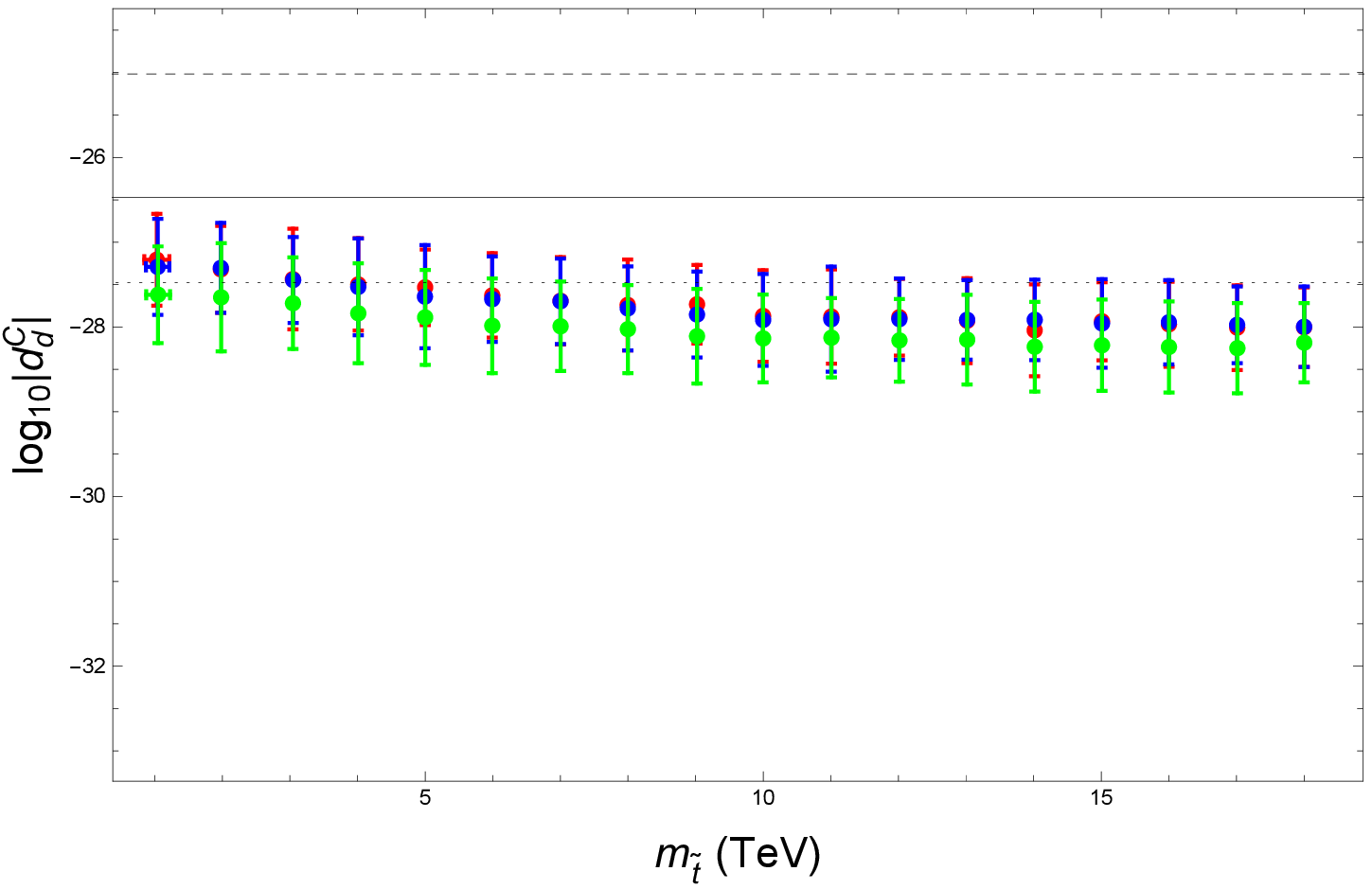,width=0.87\textwidth}
    \caption{$m_{\tilde{t}}$ dependence of $d_u^C$ (upper panel) and $d_d^C$ (lower panel). The red, blue and green plots are \textit{real $Y_u$ type}, \textit{complex $Y_u$ type} and the \textit{$E_6$ model}, respectively. Each vertical error bar shows the standard deviation for the predicted values of 
$\log {}_{10}|d_q^C|\: (q=u,d)$ which are obtained in various model points
with different $O(1)$ coefficients. The horizontal error bar shows the distribution of stop masses by variation of $O(1)$ coefficients of
    Yukawa couplings and $A$ parameters. The black solid line is current bound from Hg EDM and allowed region is lower area. The dashed line shows the
     current bound from neutron EDM, and the dotted line is the bound expected in future experiments of
neutron EDM. In these figures, we set $m_0=10$ TeV and $A_0=-1$ TeV.}
    \label{fig:stopdepduCddC}
  \end{center}
\end{figure}

\begin{figure}[t]
  \begin{center}
    \epsfig{figure=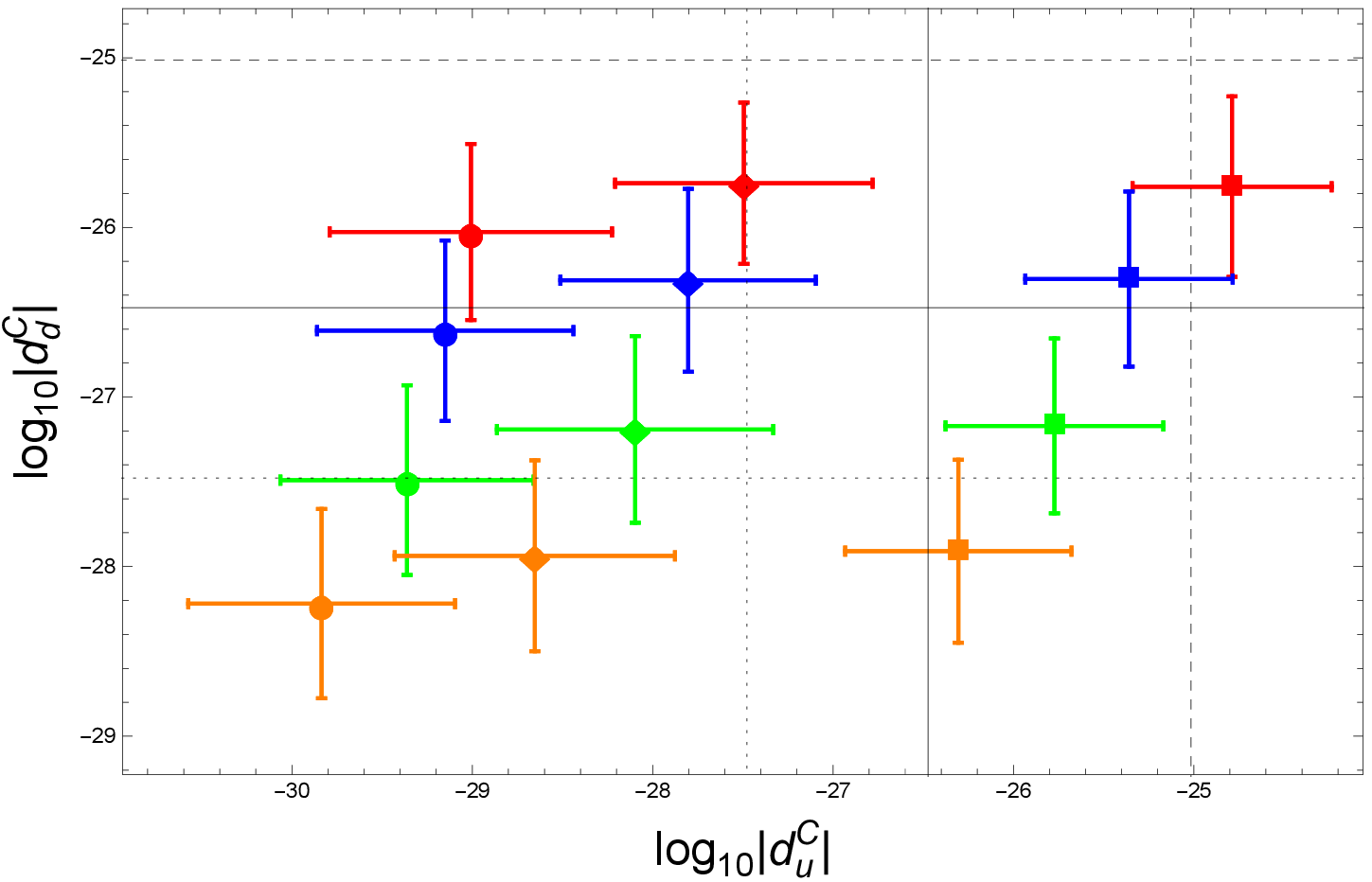,width=0.87\textwidth}\\
    \epsfig{figure=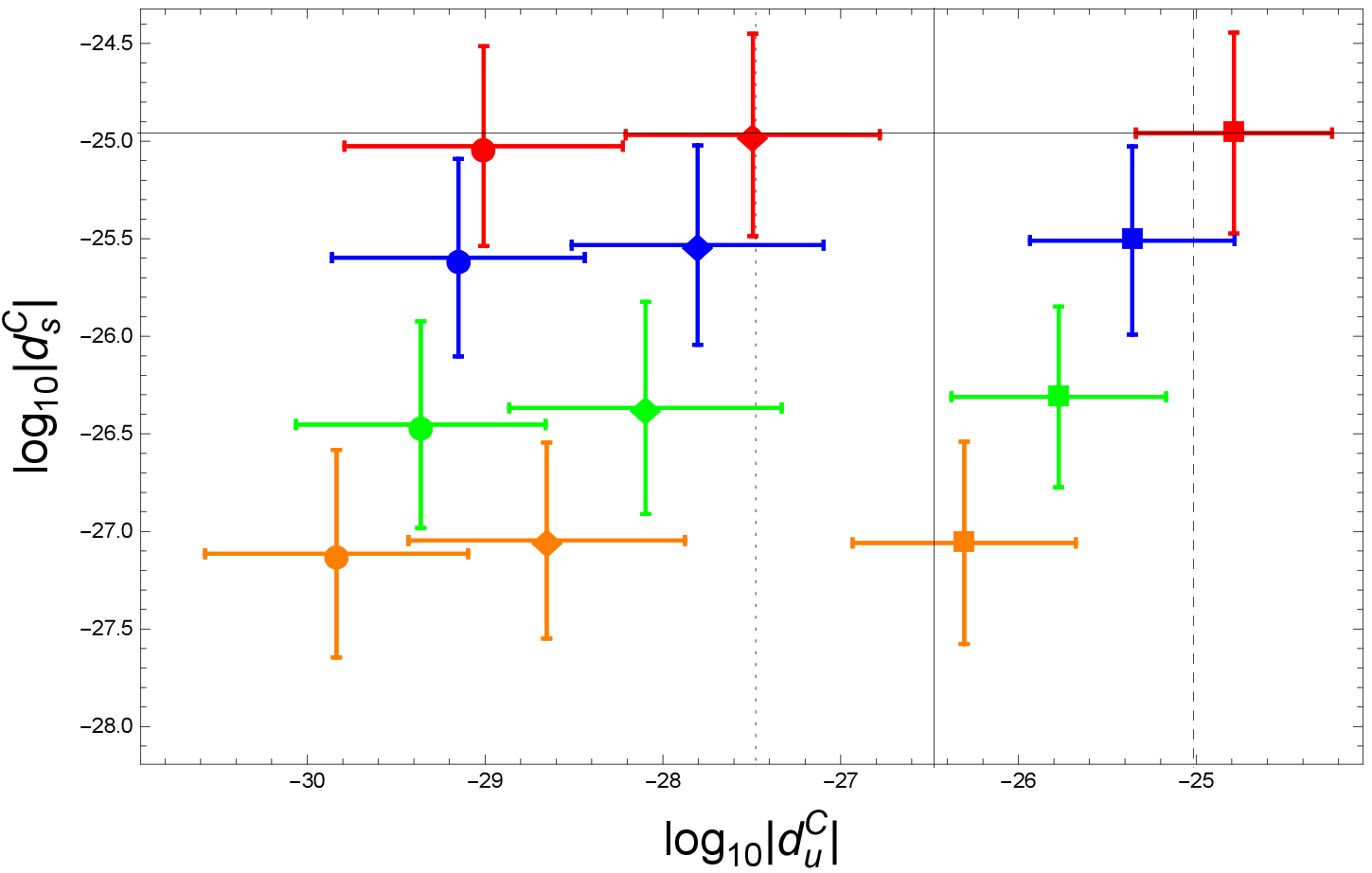,width=0.87\textwidth}
    \caption{Up, down and strange quark CEDM in three type of boundary condition of $Y_u$. Upper panel is up and down quark CEDM and lower panel is up and strange quark CEDM. Diamond, square and circle plots are \textit{real $Y_u$ type}, \textit{complex $Y_u$ type} and \textit{$E_6$ model}, respectively. Red, blue, green and orange mean that $m_0$ is 5 TeV, 10 TeV, 20 TeV and 40 TeV. Each error bar shows the standard deviation for the predicted values of 
$\log {}_{10}|d_q^C|\: (q=u,d,s)$ which are obtained in various model points
with different $O(1)$ coefficients.  Black solid line is the current bound from Hg EDM and allowed region is lower left area. Dashed line shows the
     current bound from neutron EDM, and the dotted line is the bound expected in future experiments of
neutron EDM.}
    \label{fig:quarkCEDMs}
  \end{center}
\end{figure}

We have calculated CEDMs in $O(100)$ model points with different $O(1)$
coefficients and obtained the average and the standard deviation of 
log${}_{10}|d_q^C|$  which are shown in Figs. \ref{fig:m0depduCddC}--\ref{fig:electronEDM} as the center value and the 
error bar, respectively. For simplicity, we do not impose the conditions to obtain realistic
CKM matrix in our calculation. 
The result is below. First we show $m_0 (m_{\tilde d})$ dependence of CEDMs in Fig. \ref{fig:m0depduCddC}. The vertical axis is log${}_{10}|d_u^C|$ (upper panel) and log${}_{10}|d_d^C|$ (lower panel), and the horizontal axis is heavy sfermion mass at low energy denoted as $m_{\tilde{d}}$, which is almost determined by $m_0$ value. Red, blue and green plots are \textit{real $Y_u$ type}, \textit{complex $Y_u$ type} and \textit{$E_6$ model}, respectively. 
Black solid line is current bound from Hg EDM and allowed region is lower area.
Dashed line shows the current bound from neutron EDM, and the dotted line is the bound expected in future experiments of
neutron EDM\cite{neutronFuture}. 
We set $A_0=-1$ TeV 
at the GUT scale. In these figures, we choose $M_{1/2}$ and $m_3$ so that stop mass at the SUSY scale\footnote{
In this paper, we take the SUSY scale = 1 TeV as the renormalization scale, even when the squark
masses are much larger than 1 TeV.}
 become about 2 TeV in each $m_0$ case, and $M_{\tilde{g}}$ and $|A_{u33}|$ value at the SUSY scale is shown in Table \ref{tab:Au33stopfix}.
From Fig. \ref{fig:m0depduCddC}, it is clear that $d_d^C$ is decoupled when $m_0$ increases and we found that roughly $m_{\tilde{d}} >$ 7 TeV is required to satisfy the current bound from Hg EDM, corresponding to the situation setting $m_0>7$ TeV at GUT scale. (In this paper, we discuss the limit of sfermion masses by using the center value in the distribution.) However, because of nondecoupling effect caused by stop contribution, current bound for $d_u^C$ is severe if $Y_u$ is complex at GUT scale. In order to satisfy the current bound in the \textit{complex $Y_u$ type}, $m_{\tilde{t}}$ must be larger.

\begin{table}[t]
  \begin{center}
    \begin{tabular}{|c||c|c|}
      \hline
      $m_0$ (TeV) & $M_{\tilde g}$ (TeV) & $|A_{u33}|$ (TeV) \\ \hline \hline
      5  & 2.7 & $2.0$ \\ \hline
      10 & 2.8 & $2.1$ \\ \hline
      20 & 4.3 & $3.1$ \\ \hline
      30 & 6.2 & $4.4$ \\ \hline
      40 & 8.4 & $5.8$ \\ \hline
    \end{tabular}
    \caption{$M_{\tilde{g}}$ and $|A_{u33}|$ at the SUSY scale (1 TeV) in each $m_0$ value for calculation in Fig. \ref{fig:m0depduCddC}.}
    \label{tab:Au33stopfix}
  \end{center}
\end{table}

How large $m_{\tilde t}$ is required to suppress $d_u^C$ sufficiently?
We show $m_{\tilde{t}}$ dependence of CEDMs in Fig. \ref{fig:stopdepduCddC}. The vertical axis is log${}_{10}|d_u^C|$ (upper panel) and log${}_{10}|d_d^C|$ (lower panel), and the horizontal axis is $m_{\tilde{t}}$. The colors of plots and shapes of lines have the
same meanings as in Fig. \ref{fig:m0depduCddC}.
%
In these figures, we set $m_0=20$ TeV and $A_0=-1$ TeV at GUT scale. $M_{\tilde g}=3$ TeV and 
$m_{\tilde t}\equiv \sqrt{m_{\tilde t1}m_{\tilde t2}}$ are given at the SUSY scale, where $m_{\tilde t_1}^2$ and
$m_{\tilde t_2}^2$ are eigenvalues of the matrix of stop mass square. Then $A_{u33}\sim 2.2$ TeV at the
SUSY scale.
From Fig. \ref{fig:stopdepduCddC}, it is easy to understand that $d_u^C$ is strongly dependent on $m_{\tilde{t}}$, while $d_d^C$ is almost independent.
The flat regions appear also in $d_u^C$,  which are caused by the contributions from the first two generation
squarks.
 Roughly, when $m_{\tilde{t}}>7$ TeV, current bound of $d_u^C$ from Hg EDM is satisfied even if $Y_u$ is complex at GUT scale. (Note that this lower limit for $m_{\tilde t}$ is not so far from the prediction obtained
 by the MIA as well as the lower limit for $m_{\tilde d}$. ) However, from the point of view of naturalness, it is preferable that $m_{\tilde{t}}$ has smaller value, so such a large stop mass may not be acceptable. \textit{real $Y_u$ type} and \textit{$E_6$ model} are satisfying $d_u^C$ bound even if the stop mass is smaller than 1 TeV. Therefore, real $Y_u$ at GUT scale can be an important condition to satisfy $d_u^C$ bound when $m_{\tilde{t}}\sim \mathcal{O}(1)$ TeV.



We have investigated $m_{\tilde{d}}$ or $m_{\tilde{t}}$ dependence of strange quark CEDM $d_s^C$. The results are very similar to $d_d^C$ results and constraints of $d_s^C$ are much weaker than that of $d_d^C$ (see Fig. \ref{fig:quarkCEDMs}.), and therefore, 
we do not discuss the strange quark CEDM in detail in this paper.

Finally, we check whether 125 GeV Higgs mass is really obtained in our setup. To do this, we use \verb|FeynHiggs-2.10| \cite{FeynHiggs}. We set GUT scale parameters as shown in Table \ref{tab:inputs}.
$M_{1/2}$, $m_3$, and $A_0$ are chosen so that all sfermions have positive squared masses at SUSY scale. 
We also show $m_{\tilde{t}}$ and $|A_{u33}|$ at SUSY scale in Table \ref{tab:SUSYpara}. We found that the 125 GeV Higgs mass is realized in all three types.
\begin{table}[htb]
  \begin{center}
    \begin{tabular}{|c||c|c|c|}
      \hline
      $m_0$ (TeV) & $m_3$ (TeV) & $M_{1/2}$ (TeV) & $A_0$ (TeV) \\ \hline \hline
      5  & $1.2$ & $1.5$ & $-5.0$ \\ \hline
      10 & $1.5$ & $1.8$ & $-4.5$ \\ \hline
      20 & $2.0$ & $2.4$ & $-2.5$ \\ \hline
      40 & $3.5$ & $4.5$ & $-3.5$ \\ \hline
    \end{tabular}
    \caption{GUT scale parameters which we use in each $m_0$ value.}
    \label{tab:inputs}
  \end{center}
\end{table}
\begin{table}[htb]
  \begin{center}
    \begin{tabular}{|c||c|c|}
      \hline
      $m_0$ (TeV) & $m_{\tilde{t}}$ (TeV) & $|A_{u33}|$ (TeV) \\ \hline \hline
      5  & 1.9 & 3.4 \\ \hline
      10 & 2.3 & 3.7 \\ \hline
      20 & 2.6 & 4.1 \\ \hline
      40 & 4.3 & 7.4 \\ \hline
    \end{tabular}
    \caption{
      $m_{\tilde{t}}$ and $|A_{u33}|$ at SUSY scale in each $m_0$ value.}
    \label{tab:SUSYpara}
  \end{center}
\end{table}
The values of CEDMs in each situation are shown in Fig. \ref{fig:quarkCEDMs}. Upper panel is $d_u^C$ versus $d_d^C$, and lower panel is $d_u^C$ versus $d_s^C$. In these figures, diamond, square and circle plots are corresponding to
\textit{real $Y_u$ type}, \textit{complex $Y_u$ type} and \textit{$E_6$ model}, respectively. Red, blue, green and orange means that $m_0$ is 5 TeV, 10 TeV, 20 TeV and 40 TeV. Each error bar shows the standard deviation for the predicted values of 
$\log {}_{10}|d_q^C|\: (q=u,d,s)$ which are obtained in various model points
with different $O(1)$ coefficients. Black solid line is current bound from Hg EDM and allowed region is lower left area. Dashed line shows the
     current bound from neutron EDM, and the dotted line is the bound expected in future experiments of
neutron EDM.
From Fig. \ref{fig:quarkCEDMs}, it is clear that $d_u^C$ bound for \textit{complex $Y_u$ type} is severe, even when 125 GeV Higgs mass is realized. In \textit{$E_6$ model}, each CEDM value is smaller than that of the other two types because of the special structures of $Y_u$ and $A_u$ at the GUT scale. Therefore, these structures have some effects to suppress $|d_u^C|$ value.

\section{COMMENT ON ELECTRON EDM}

Recently, the constraint of electron EDM, $d_e$, is improved \cite{eEDMcurrent} and may be severe for this discussion. Then we also check the constraint of $d_e$ in the same situations discussed above. Note that we evaluate $d_e$ by using the expression based on Ref. \cite{CEDMbounds}. Although there are mainly two types of contributions to $d_e$ in SUSY model, neutralino and chargino contributions, we ignore the chargino contributions because Higgsinos are heavier than wino.
So, we will show the $d_e$ result for the sum of four neutralino contributions in our setup.

\begin{figure}[t]
  \begin{center}
    \epsfig{figure=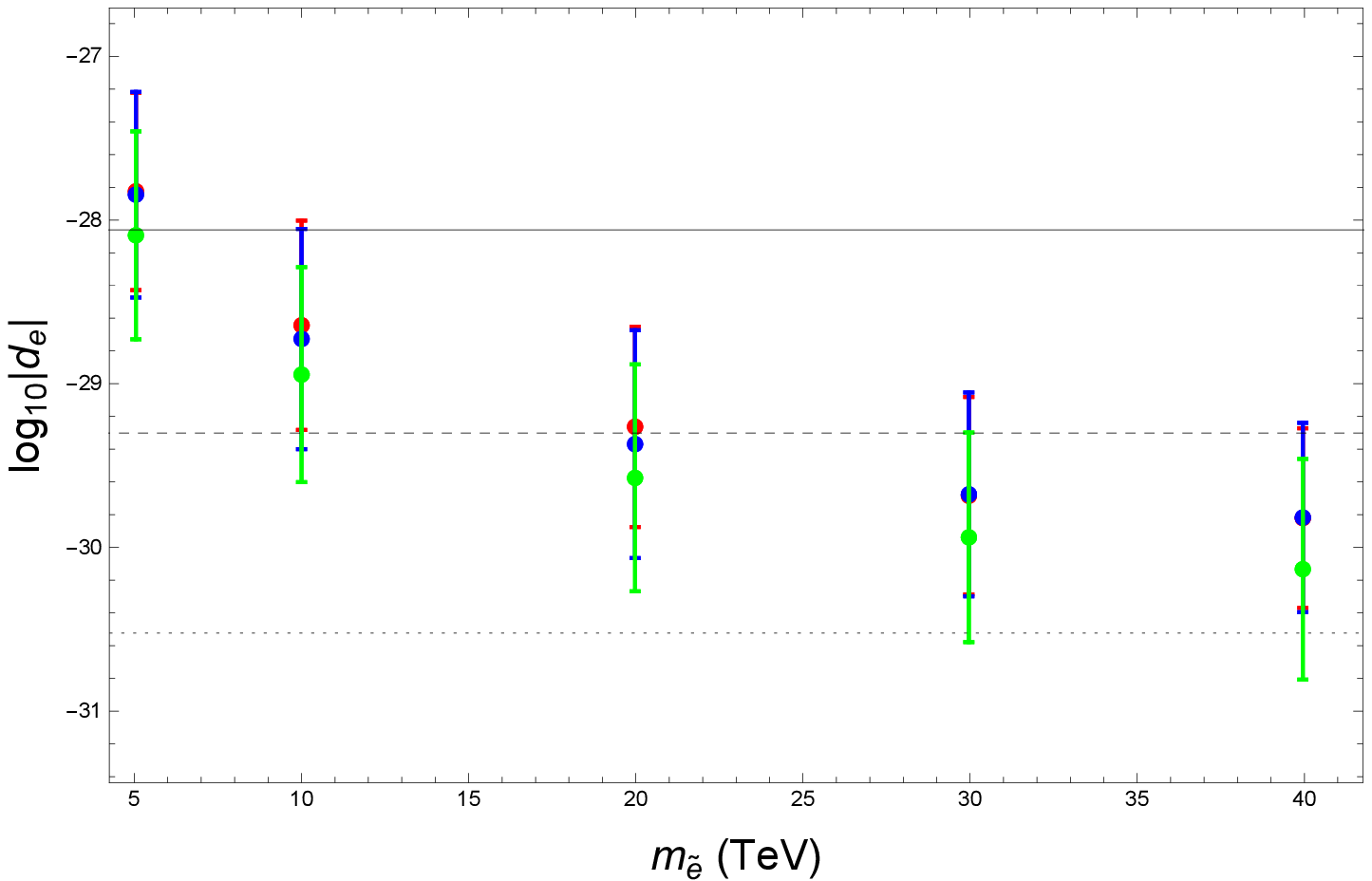,width=0.87\textwidth}
    \caption{$m_{\tilde{e}}$ dependence of $d_e$ in three type of boundary condition of $Y_u$. Red, Blue and Green plots are \textit{real $Y_u$ type}, \textit{complex $Y_u$ type} and \textit{$E_6$ model}, respectively. 
Each error bar shows the standard deviation for the predicted values of 
$\log {}_{10}|d_e|$ which are obtained in various model points
with different $O(1)$ coefficients.   
Black solid line is current bound and allowed region is the lower area. Other input parameters are same as for the Fig. \ref{fig:m0depduCddC}. The dashed(dotted) line shows the future bound
    expected by ACME II (III)\cite{eEDMfuture}. }
    \label{fig:electronEDM}
  \end{center}
\end{figure}
The result is shown in Fig. \ref{fig:electronEDM}. The vertical axis is log${}_{10}|d_e|$. In this case, we set horizontal axis as slepton mass at low energy, which is almost determined by $m_0$ value. Red, blue and green plots are the \textit{real $Y_u$ type}, \textit{complex $Y_u$ type} and \textit{$E_6$ model}, respectively. 
Black solid line is current bound, $|d_e| < 8.7\times 10^{-29}\: e\: {\rm cm}$ \cite{eEDMcurrent}, and allowed region is lower area. Other input parameters which are used for calculation are the same as for the Fig. \ref{fig:m0depduCddC}. Neutralino masses at SUSY scale in each $m_0$ case are shown in Table \ref{tab:nmass}.
\begin{table}[htb]
  \begin{center}
    \begin{tabular}{|c||c|c|c|}
      \hline
      $m_0$ (TeV) & $m_{N_1}$ (TeV) & $m_{N_2}$ (TeV) & $m_{N_3}$, $m_{N_4}$ (TeV) \\ \hline \hline
      5  & 0.5 & 0.9 & 2.1 \\ \hline
      10 & 0.6 & 1.0 & 2.4 \\ \hline
      20 & 0.8 & 1.5 & 3.3 \\ \hline
      30 & 1.2 & 2.2 & 4.2 \\ \hline
      40 & 1.6 & 3.0 & 5.2 \\ \hline
    \end{tabular}
    \caption{Neutralino masses at SUSY scale in each $m_0$ case. In this calculation, the masses of two heavy neutralinos, $m_{N_3}$ and $m_{N_4}$, are almost degenerated.}
    \label{tab:nmass}
  \end{center}
\end{table}

Compared with Figs. \ref{fig:m0depduCddC} and \ref{fig:electronEDM}, we found that constraint of $d_e$ is weaker than that of $d_d^C$ at least
in the situation we discussed in this paper. 
Roughly speaking, $m_{\tilde{e}} >$ 6 or 7 TeV is required for the $d_e$ bound in \textit{real $Y_u$ type} and \textit{complex $Y_u$ type} while $d_e$ bound is still satisfied with $m_{\tilde{e}} >$ 5 TeV in the \textit{$E_6$ model}. This is because there are special structures not only for $Y_u$ and $Y_d$ but also for the $Y_e$ in the \textit{$E_6$ model}. Electron EDM experiments will be improved in a few years \cite{eEDMfuture}, so we should care about not only the CEDM bounds but also this bound.
The expected future bounds are presented in Fig. \ref{fig:electronEDM}. Interestingly, we can expect a signal of the electron EDM in the future. If it is not seen,
$m_0$ must be larger than 40 TeV.

\section{SUMMARY AND DISCUSSION}

In this paper, we discussed the CEDM bounds in the SUSY model with the natural SUSY-type sfermion mass spectrum in which the stop masses $m_{\tilde t}$ are 
$O(1)$ TeV while the other squark masses $m_0$ are much larger than
$m_{\tilde t}$ since the CEDM constraints, especially from the Hg EDM, give severe
constraint to this natural SUSY-type sfermion mass spectrum even if
real SUSY-breaking parameters are assumed.
We calculated the CEDM of up, down, and strange quarks numerically 
at the three boundary conditions for Yukawa couplings at the GUT scale, the \textit{real $Y_u$ type}, \textit{complex $Y_u$ type}, and the \textit{$E_6$ model},
and discussed their decoupling features.
First, we concluded that the up-quark CEDM becomes sufficiently small to
 satisfy the experimental bound when up-quark Yukawa couplings are real at
 the GUT scale even if we take $m_{\tilde t} \sim O(1)$ TeV not to destabilize
the weak scale, while it becomes too large when the Yukawa couplings are
complex even if $m_0\gg m_{\tilde t}$. 
On the other hand, the down and strange quark CEDM 
become sufficiently small if $m_0>7$ TeV because of the decoupling feature.
Second, to satisfy the up-quark CEDM constraint with complex $Y_u$,  $m_{\tilde t}>7$ TeV is needed, which destabilizes the weak scale.



In the natural SUSY-type sfermion mass spectrum, off-diagonal elements of $(\Delta^d_{RR})$, which is defined as the same rule in Eq.~(\ref{eq:66sfmass}) are strongly suppressed because the masses of the right-handed sdown type are almost degenerated. For this reason, a dominant contribution to $d_d^C$ is proportional to Im$\left[ \left(\Delta^d_{LL}\right)_{31} \left(\Delta^d_{RL}\right)_{13} \right]$ ($\propto A_{d13}$ in the super-CKM basis) rather than Im$\left[ \left(\Delta^d_{LL}\right)_{31} \left(\Delta^d_{RL}\right)_{33} \left(\Delta^d_{RR}\right)_{13} \right]$. When the $A$ parameters are proportional to the corresponding Yukawa couplings $A_f \propto Y_f$~$(f = u, d, e)$ at the GUT scale, off-diagonal elements of $A$ parameters in the super-CKM basis are suppressed at the SUSY-breaking scale and, therefore, $d_d^C$ is strongly suppressed ($\sim \mathcal{O}(10^{-32})\ e$ cm). In such a case, we cannot constrain the sfermion masses from the bound of $d_d^C$ even when the CEDM constraints are improved at the level of the future neutron EDM sensitivity. Note that this situation can be also realized when $A_0 = 0$ at the GUT scale. On the other hand, in the case of $A_f \propto Y_f$ at the GUT scale, $d_u^C$ does not change so much in the natural SUSY-type sfermion mass spectrum because the dominant contribution to $d_u^C$ is proportional to the diagonal element of the $A$ parameter $A_{u33}$ in the super-CKM basis as discussed in Eq.~(\ref{eq:duCapp}). The situation does not change so much in the case of $A_0 = 0$ at the GUT scale since the $A_0$ value does not affect the value of the diagonal elements of the $A$ parameters at the SUSY-breaking scale so much because of the RGE running. We checked these behaviors numerically, and the lower bounds of $m_{\tilde{t}}$ from the bound of $d_u^C$ are the same as our results.

These constraints are dependent on explicit models for Yukawa couplings and the sfermion mass spectrum. 
In this paper, we have just demonstrated the constraints from the EDMs in a certain model for the Yukawa 
couplings and the sfermion mass spectrum, which are obtained from the $E_6$ GUT with family symmetry. 
Therefore, the constraints become
different from ours if different models for Yukawa couplings 
are adopted.
However, we note that 
our model will give comparatively weaker constraints than the others because the diagonalizing matrices
of the up-type Yukawa matrix have small mixings. (Of course, we can consider
the models which give weaker constraints than ours, for example, the diagonal up-type Yukawa matrix model.) 
Therefore, our constraints to the natural SUSY-type sfermion mass models from $d_u^C$ are quite general.

In this paper, we have neglected the uncertainties in the calculation of the relation between the Hg (neutron) EDM and CEDMs and
discussed the constraints. However, the uncertainties for the coefficients are more than 100 \% for the Hg EDM.
Conservatively, we may have to use the constraints only from the neutron EDM.
Then, we have almost no constraints from the neutron EDM to the natural SUSY model with $O(1)$ TeV stop mass.
In that case, constraints from the electron EDM become important and give lower bounds of $m_0$,
although no constraint for $m_{\tilde t}$ is given.
Since the experimental sensitivity of the electron EDM is expected to be improved by about 2 orders of magnitude, we can expect that 
nonvanishing electron EDM is observed in future experiments.
If it is not observed, 
the $m_0$ is expected to be larger than 40 TeV. 
To improve the bound for $m_{\tilde t}$, future experiments of neutron EDM are important. Since experimental sensitivity
of the neutron EDM is expected to be improved by more than two orders of magnitude, the observation of nonvanishing EDM
of neutron is expected. No signal means $m_0>20$ TeV, and $m_{\tilde t}>20$ TeV if $Y_u$ is complex,
while almost no constraint to $m_{\tilde t}$ if $Y_u$ is real. One more way to improve
the bound for $m_{\tilde t}$ is, of course,  to reduce the uncertainties in theoretical calculation of Hg EDM.


We conclude that the up quark CEDM constraint can be severe in natural 
SUSY type sfermion mass spectrum. 
If experimental bounds of EDM and/or theoretical calculation of Hg EDM are improved in future, they will constrain the lower bounds of stop mass and the other heavy sfermion masses. In addition, if the sfermion masses, especially the stop mass, are observed in future experiments, we may be able to constrain the structure of $Y_u$ at GUT scale by the CEDM constraints.

\section*{ACKNOWLEDGMENTS}
We thank F. Sala and J. Hisano for advice on the present situation of theoretical calculation of Hg EDM.
N. M. is supported in part by Grants-in-Aid for Scientific Research from MEXT of Japan (No. 15K05048).
Y.M. is supported in part by 
the National Research Foundation of Korea (NRF) Research Grant No.
NRF- 2015R1A2A1A05001869 and 
the National Natural Science Foundation of China (NSFC) under Contracts 
No. 11435003, No. 11225523, and No. 11521064.
The work of Y.S. is supported by the Japan Society for the Promotion of Science (JSPS) Research Fellowships for Young Scientists, No. 16J08299.


\section*{APPENDIX: LOOP INTEGRAL FOR THE DIAGRAM IN FIG. \ref{fig:CEDMdiagram}(b)}
The expression of the up-quark CEDM $d_u^C$ in the mass insertion approximation is
\begin{eqnarray}
  d_u^C &=& \frac{\alpha_s}{4\pi} \frac{M_{\tilde{g}}}{m_{\tilde{t}}^2} {\rm Im}\left[(\delta^u_{LL})_{31} (\delta^u_{RL})_{33} (\delta^u_{RR})_{13}\right] \times F_{\rm MIA}(r_{\tilde{g}},r_{\tilde{t}}) \\[0.5ex]
  F_{\rm MIA}(r_{\tilde{g}},r_{\tilde{t}}) &\equiv& 6 \, r_{\tilde{t}}^2 \left(-3 I_G(r_{\tilde{g}},r_{\tilde{t}}) + \frac{1}{3} I_{S_1}(r_{\tilde{g}},r_{\tilde{t}}) + \frac{1}{3} I_{S_2}(r_{\tilde{g}},r_{\tilde{t}}) + \frac{1}{3} I_{S_3}(r_{\tilde{g}},r_{\tilde{t}}) + \frac{1}{3} I_{S_4}(r_{\tilde{g}},r_{\tilde{t}})\right) \nonumber \\
  \label{eq:MIAduC}
\end{eqnarray}
where $r_{\tilde{g}}=\frac{M_{\tilde{g}}^2}{m_{\tilde{u}}^2}$, $r_{\tilde{t}}=\frac{m_{\tilde{t}}^2}{m_{\tilde{u}}^2}$ and $I_i(r_{\tilde{g}},r_{\tilde{t}})$ are loop integrals. Each integral is
\begin{eqnarray}
  I_G(r_{\tilde{g}},r_{\tilde{t}}) &=& \int_0^1 dx_1\cdots dx_4 \delta(\Sigma_ix_i - 1) \frac{2 x_1 x_3 x_4}{[r_{\tilde{g}} (x_1+x_2) + x_3 + r_{\tilde{t}} x_4]^4}, \\
  I_{S_1}(r_{\tilde{g}},r_{\tilde{t}}) &=& \int_0^1 dx_1\cdots dx_4 \delta(\Sigma_ix_i - 1) \frac{(2 x_3 + 2 x_4 - 1) x_3 x_4}{[r_{\tilde{g}} x_1 + x_2 + x_3 + r_{\tilde{t}} x_4]^4}, \\
  I_{S_2}(r_{\tilde{g}},r_{\tilde{t}}) &=& \int_0^1 dx_1\cdots dx_5 \delta(\Sigma_ix_i - 1) \frac{(2 x_3 + 2 x_5 - 1) x_5}{[r_{\tilde{g}} x_1 + x_2 + x_3 + r_{\tilde{t}} (x_4 + x_5)]^4}, \\
  I_{S_3}(r_{\tilde{g}},r_{\tilde{t}}) &=& \int_0^1 dx_1\cdots dx_5 \delta(\Sigma_ix_i - 1) \frac{(2 x_3 + 2 x_5 - 1) x_4}{[r_{\tilde{g}} x_1 + x_2 + x_3 + r_{\tilde{t}} (x_4 + x_5)]^4}, \\
  I_{S_4}(r_{\tilde{g}},r_{\tilde{t}}) &=& \int_0^1 dx_1\cdots dx_4 \delta(\Sigma_ix_i - 1) \frac{(2 x_3 - 1) x_2 x_4}{[r_{\tilde{g}} x_1 + x_2 + x_3 + r_{\tilde{t}} x_4]^4}.
\end{eqnarray}

We show the values of $F_{\rm MIA}(r_{\tilde{g}},r_{\tilde{t}})$ with several values of mass ratio, $r_{\tilde{g}}$ and $r_{\tilde{t}}$ (see Table \ref{LF}).
\begin{table}[h]
  \begin{center}
    \begin{tabular}{|c||c|c|c|c|c|c|}
      \hline
      $r_{\tilde{t}}$ \verb|\| $r_{\tilde{g}}$ & $0.2^2$ & $0.3^2$ & $0.5^2$ & $1^2$ & $2^2$ & $5^2$ \\ \hline\hline
      $0.1^2$ & $4.1 \times 10^{-2}$ & $1.1 \times 10^{-2}$ & $1.8 \times 10^{-3}$ & $1.2 \times 10^{-4}$ & $6.9 \times 10^{-6}$ & $1.6 \times 10^{-7}$ \\ \hline
      $0.2^2$ & $2.4 \times 10^{-1}$ & $7.9 \times 10^{-2}$ & $1.6 \times 10^{-2}$ & $1.2 \times 10^{-3}$ & $6.9 \times 10^{-5}$ & $1.5 \times 10^{-6}$ \\ \hline
      $0.3^2$ & $5.3 \times 10^{-1}$ & $2.0 \times 10^{-1}$ & $4.6 \times 10^{-2}$ & $4.0 \times 10^{-3}$ & $2.4 \times 10^{-4}$ & $5.1 \times 10^{-6}$ \\ \hline
      $0.4^2$ & $8.5 \times 10^{-1}$ & $3.6 \times 10^{-1}$ & $9.1 \times 10^{-2}$ & $8.8 \times 10^{-3}$ & $5.6 \times 10^{-4}$ & $1.2 \times 10^{-5}$ \\ \hline
      $0.5^2$ & $1.2$               & $5.2 \times 10^{-1}$ & $1.4 \times 10^{-1}$ & $1.5 \times 10^{-2}$ & $1.0 \times 10^{-3}$ & $2.1 \times 10^{-5}$ \\ \hline
      $0.6^2$ & $1.4$               & $6.8 \times 10^{-1}$ & $2.0 \times 10^{-1}$ & $2.3 \times 10^{-2}$ & $1.6 \times 10^{-3}$ & $3.4 \times 10^{-5}$ \\ \hline
      $0.7^2$ & $1.7$               & $8.2 \times 10^{-1}$ & $2.6 \times 10^{-1}$ & $3.2 \times 10^{-2}$ & $2.3 \times 10^{-3}$ & $5.0 \times 10^{-5}$ \\ \hline
      $0.8^2$ & $1.9$               & $9.6 \times 10^{-1}$ & $3.2 \times 10^{-1}$ & $4.1 \times 10^{-2}$ & $3.2 \times 10^{-3}$ & $6.8 \times 10^{-5}$ \\ \hline
      $0.9^2$ & $2.1$               & $1.1$               & $3.7 \times 10^{-1}$ & $5.1 \times 10^{-2}$ & $4.1 \times 10^{-3}$ & $8.9 \times 10^{-5}$ \\ \hline
      $1^2$   & $2.3$               & $1.2$               & $4.2 \times 10^{-1}$ & $6.1 \times 10^{-2}$ & $5.0 \times 10^{-3}$ & $1.1 \times 10^{-4}$ \\ \hline
      $2^2$   & $3.2$               & $1.8$               & $7.6 \times 10^{-1}$ & $1.4 \times 10^{-1}$ & $1.5 \times 10^{-2}$ & $4.0 \times 10^{-4}$ \\ \hline
      $5^2$   & $3.7$               & $2.3$               & $1.0$               & $2.4 \times 10^{-1}$ & $3.3 \times 10^{-2}$ & $1.2 \times 10^{-3}$ \\ \hline
    \end{tabular}
    \caption{The values of $F_{\rm MIA}(r_{\tilde{g}},r_{\tilde{t}})$ with several values of $r_{\tilde{g}}$ and $r_{\tilde{t}}$.}
    \label{LF}
  \end{center}
\end{table}

\end{document}